\documentclass[a4paper,aps,reprint,superscriptaddress]{revtex4-1}

\usepackage[top=2.3cm, bottom=2.2cm, left=1.8cm, right=1.8cm]{geometry}

\usepackage{amsfonts,amsmath,amssymb,amsfonts,latexsym}

\usepackage[utf8]{inputenc}
\usepackage{color}
\usepackage{graphicx}

\usepackage{xcolor}
\definecolor{myurlcolor}{HTML}{08457E}
\definecolor{mylinkcolor}{HTML}{2A52BE}
\definecolor{mycitecolor}{HTML}{E30022}

\usepackage{float}
\usepackage[colorlinks, linkcolor=mylinkcolor, citecolor=mycitecolor, urlcolor=myurlcolor, linktocpage=true]{hyperref}

\def\equationautorefname~#1\null{(#1)\null}
\def\tableautorefname~#1\null{(#1)\null}
\def\figureautorefname~#1\null{(#1)\null}
\def\sectionautorefname~#1\null{(#1)\null}

\def\H{{\cal H}}      
\def\K{{\cal K}}      

\newcommand{\sfrac}[2]{\dfrac{\,#1\,}{\,#2\,}}

\newcommand{\der}[2]{\sfrac{\partial #1}{\partial #2}}

\newcommand{\dder}[2]{\sfrac{\partial^{\,2} #1}{\partial #2^2}}

\let\oldsqrt\sqrt
\def\sqrt{\mathpalette\DHLhksqrt}
\def\DHLhksqrt#1#2{%
    \setbox0=\hbox{$#1\oldsqrt{#2\:}$}\dimen0=\ht0
    \advance\dimen0-0.4\ht0
    \setbox2=\hbox{\vrule height\ht0 depth -\dimen0}%
    {\box0\lower0.4pt\box2}\,}

\newcommand{\dl}[1]{\partial_{#1}}
\newcommand{\mf}[1]{\mathbf{#1}}

\newcommand{\al}{\alpha}
\newcommand{\bt}{\beta}

\newcommand{\gm}{\gamma}

\newcommand{\be}[1]{\begin{equation}\label{#1}}
\newcommand{\ee}{\end{equation}}
\newcommand{\ba}[1]{\begin{eqnarray}\label{#1}}
\newcommand{\ea}{\end{eqnarray}}
\newcommand{\rf}[1]{(\ref{#1})}
\newcommand{\nn}{\nonumber}

\begin{document}

\title{Effect of the spatial curvature of the Universe \\on the form of the gravitational potential}

\author{Maxim Eingorn}
\email{maxim.eingorn@gmail.com}
\affiliation{Department of Mathematics and Physics, North Carolina Central University, Fayetteville st. 1801, Durham, North Carolina 27707, U.S.A.}

\author{A. Emrah Y\"{u}kselci}
\email{yukselcia@itu.edu.tr}
\affiliation{Department of Physics, Istanbul Technical University, 34469 Maslak, Istanbul, Turkey}

\author{Alexander Zhuk}
\email{ai.zhuk2@gmail.com}
\affiliation{Department of Physics, Istanbul Technical University, 34469 Maslak, Istanbul, Turkey}
\affiliation{Astronomical Observatory, Odessa National University,
Dvoryanskaya st. 2, 65082 Odessa, Ukraine}

\date{\today}

\begin{abstract}
Within the cosmic screening approach, we obtain the exact formulas for the velocity-independent gravitational potentials produced by matter in the form of discrete sources distributed in the open and closed Universes. These formulas demonstrate that spatial curvature of the Universe considerably affect the form of
the potentials and forces. While in the open Universe the gravitational force undergoes exponential suppression at cosmological distances, in the closed Universe the force induced by an individual mass is equal to zero at the antipodal point with respect to this mass. The derived formulas are applicable for investigations of the motion of astrophysical objects (e.g., galaxies) in the open and closed Universes, and for simulations of the large scale structure formation.
\end{abstract}

\maketitle

\vspace{.5cm}



\raggedbottom

\section{Introduction}

\setcounter{equation}{0}

Following the natural assumption that laws of physics should be the same wherever in the Universe, we arrive at the conclusion that at sufficiently large scales
our Universe should be homogeneous and isotropic. This statement is known as the cosmological principle \cite{Weinberg}. It is well known that such a homogeneous
and isotropic space is the constant curvature space with three possible cases for the spatial metric: constant positive curvature (closed Universe), constant
negative curvature (open Universe) and zero curvature (flat Universe) \cite{MTW}.


Within the appropriate extension of the standard $\Lambda$CDM model, the most recent analysis of the CMB data results in the spatial curvature parameter
$\Omega_K=-0.044^{+0.018}_{-0.015}$ \cite{Planck2018}. Inclusion of the lensing and BAO in the analysis gives $\Omega_K=0.0007\pm 0.0019$ \cite{Planck2018}.
Non-CMB data mildly favor a closed spatial hypersurface \cite{Ratra1}. If one uses inflation power spectra in non-flat models, there is also some evidence for a mildly closed Universe in these data \cite{Ratra2,Ratra3,Ratra4}. 
Most of such constraints are based on some cosmological models (e.g., $\Lambda$CDM), i.e. they are model-dependent. Keeping in mind the importance of the
curvature parameter (affecting, e.g., the global dynamics of the Universe, the lensing \cite{Bernstein}, the shape of the gravitational potential (see below),
etc.), it is of great interest to determine $\Omega_K$ in the model-independent way. There is extensive literature on this subject (see, e.g., the latest articles
\cite{curvature1,curvature2} and numerous references therein). According to these investigations, ``the nonzero $\Omega_K$ cannot be ruled out by the current
observations'' \cite{curvature1}. Hence, it is reasonable to study cosmological models with any sign of $\Omega_K$ (including the zero value).

It is quite expected that cosmological models with different signs of spatial curvature will lead to different physical effects. For example, the dynamics of
astrophysical objects may differ in spaces with different global topologies. To study the motion of astrophysical objects in the Universe, we should know the form
of the gravitational potential created by massive inhomogeneities (e.g., stars, galaxies and groups of galaxies). It is well known that the gravitational
potential is defined by the first-order scalar perturbation of the $g_{00}$ metric component \cite{Landau,Gorbunov:2011zzc}. This perturbation satisfies the
equation which includes the curvature parameter (see, e.g., \cite{EZflow,EZremarks}). This means that the gravitational potential must have different forms for
the closed, open and flat Universe cases. We have already investigated this problem in the papers \cite{EZflow,Burgazli}. Here, we have revealed the gravitational
potential screening effect due to the presence of both the spatial curvature and an additional perfect fluid with the constant parameter $\omega=-1/3$ in the
linear equation of state. However, the cosmic screening due to the matter (both dark and baryonic) was not taken into account. In the case of the flat Universe,
the cosmic screening was thoroughly studied in the papers \cite{Eingorn:2015hza,Eingorn:2016kdt,Eingorn:2015yra,cosmlaw,Eingorn:2017adg}. Here, the analytic
expression for the gravitational potential was obtained. It was shown that the potential undergoes the Yukawa-type exponential screening at cosmological
distances. The cosmological background consisting of the average mass density of dark and baryonic matter is responsible for this screening. In the present paper
we investigate how the nonzero spatial curvature of the Universe affects this result. We find analytic expressions for the gravitational potentials for each type
of curvature and demonstrate that the form and properties of the potential considerably depend on the curvature type choice.

The paper is structured as follows. In Sec. 2, we describe the model and present the equation for the gravitational potential within the cosmic screening
approach. The general solution of this equation is given in Sec. 3. In Sections 4, 5 and 6, we analyze the potentials for the flat, open and closed Universe
cases, respectively. The main results are briefly summarized in concluding Sec. 7.

\section{Setting of the problem}
\label{Sec2}

\setcounter{equation}{0}

We consider the homogeneous and isotropic Universe which is described by Friedmann-Lema$\mathrm{\hat{\i}}$tre-Robertson-Walker (FLRW) metric
\ba{2.1}
ds^2 &=& a^2(\eta) \big[ d\eta^2 - \gm_{\al \bt} \, dx^\al dx^\bt \,\big] \nn\\
&=& a^2(\eta) \big[ d\eta^2 - d\chi^2 - \Sigma^2(\chi)  \,d\Omega^2 \big]\, , \ea
where $a(\eta)$ is the scale factor and $\eta$ is the conformal time connected with the synchronous time $t$ as follows: $d\eta = cdt/ a$. Since the scale factor has the dimension of length, the conformal time is dimensionless. The choice of the metric in the form \rf{2.1} turns out to be convenient when constructing the perturbation theory within the cosmic screening scheme \cite{Eingorn:2015hza}. The function $\Sigma(\chi)$ is defined as
\be{2.2}
\Sigma(\chi) =
\begin{cases}
	\sin\!\chi, &\quad \chi \in [0,\pi] \;\text{for}\; \K=+1 \\
	\chi, &\quad \chi \in [0,+\infty) \;\text{for}\; \K=0 \\
	\sinh\!\chi, &\quad \chi \in [0,+\infty) \;\text{for}\; \K=-1 \\
\end{cases}
\ee where $\K=-1,0,+1$ indicates open, flat and closed Universe cases, respectively.

The Friedmann equation for the background containing nonrelativistic pressureless matter and the cosmological constant is
\be{2.3}
\sfrac{3(\H^2 + \K)}{a^2} = \kappa \bar{\varepsilon} + \Lambda\, ,
\ee
where the dimensionless parameter $\H\equiv (da/d\eta)/a =(a/c)H$, with $H\equiv (da/dt)/a$ being the Hubble parameter, $\Lambda$ is the cosmological constant, $\bar\varepsilon = \bar{\rho}c^2/a^3$ denotes the energy density of pressureless
matter with comoving mass density $\bar\rho=\mathrm{const}$, $c$ is the speed of light and overline implies the average value. Additionally, we define $\kappa \equiv 8\pi G_{\!N}/c^4$, where $G_{\!N}$ is the gravitational constant.

The cosmological parameters are defined as
\be{2.4}
\Omega_{\rm M} \equiv \sfrac{\kappa \bar{\rho} c^4}{3 H_0^2 a_0^3} \;, \quad \Omega_\Lambda \equiv \sfrac{\Lambda c^2}{3 H_0^2} \;, \quad
\Omega_\K \equiv -\sfrac{\K c^2}{a_0^2 H_0^2}\, ,
\ee
where $a_0$ and $H_0$ denote the present values of the scale factor and the Hubble parameter, respectively. For the illustrative purposes we will use the values
\ba{2.5}
\Omega_{\rm M} &=& 0.315 \;, \quad \Omega_{\K=-1} = 0.0007 \;, \quad \Omega_{\K=+1} = -0.044\, , \nn\\
H_0 &=& 67.4 \;{\rm km\, s^{-1} Mpc^{-1}}
\ea
in accord with the results of \cite{Planck2018}.

We consider matter (e.g., galaxies) in the form of discrete point-like masses with comoving mass density
\be{2.6}
\rho = \sum_n \rho_n = \sfrac{1}{\sqrt{\gm}} \sum_n m_n \, \delta(\mf{r}-\mf{r}_n) \, ,
\ee
where $\gm$ is the determinant of $\gm_{\alpha\beta}$. These discrete inhomogeneities perturb the background metric \rf{2.1}:
\be{2.7} ds^2 = a^2 \big[ (1+2\Phi) d\eta^2  - (1-2\Phi) \gm_{\al\bt} \, dx^\al dx^\bt\,\big]\, , \ee
where we restrict ourselves to scalar perturbations. The fluctuation of the energy density is given by the formula \cite{EZflow,EZremarks,Eingorn:2015hza}
\be{2.8} \delta\varepsilon= \frac{c^2\delta\rho}{a^3} + \frac{3\bar\rho c^2 \Phi}{a^3}\, , \ee
where $\delta\rho\equiv\rho-\bar\rho$. It is well known that the first-order scalar perturbation $\Phi(\eta,\mf{r})$ defines the gravitational potential
\cite{Landau,Gorbunov:2011zzc}. In our case, it is the potential created by all masses in the point $\mf{r}=\left(x^1,x^2,x^3\right)$.

Within the cosmic screening approach, the gravitational potential satisfies the following equation{\footnote{It is worth noting that we work in the weak field
		limit where the peculiar velocities are much less than the speed of light. In this case, as was shown in \cite{Eingorn:2015hza}, the peculiar velocities
		negligibly contribute to the gravitational potential. For this reason we do not include the velocity-dependent term into Eq.~\rf{2.9}.}} \cite{Eingorn:2015hza}:
\be{2.9}
\Delta \Phi + 3 \bigg(\K - \sfrac{\kappa\Bar{\rho} c^2}{2a} \bigg)\Phi = \sfrac{\kappa c^2}{2a} \delta\rho\, ,
\ee
where  the Laplace operator
\be{2.10}
\Delta = \sfrac{1}{\sqrt{\gamma}} \dl{\alpha} \big(\sqrt{\gamma}\,\gamma^{\alpha\beta}\dl{\beta}\big) \, .
\ee
As one can see, Eq.~\rf{2.9} is the Helmholtz-type equation (not the Poisson one!). The nonzero spatial curvature ($\mathcal{K}\neq0$) and background matter
density ($\bar\rho\neq0$) are responsible for this effect. The flat Universe case $\mathcal{K}=0$ has been already investigated in \cite{Eingorn:2015hza} where
the effect of the Yukawa screening of the gravitational potential at cosmological scales has been clearly demonstrated. Now we want to understand how nonzero
spatial curvature affects the shape of the gravitational potential.

\section{General solution}
\label{Sec3}

\setcounter{equation}{0}

To solve Eq.~\rf{2.9}, it is convenient to introduce a new function
\be{3.1} \varphi(\eta,\mf{r})=c^2\,a(\eta)\,\Phi(\eta,\mf{r})\, . \ee
Then Eq.~\rf{2.9} reads
\be{3.2}
\Delta\varphi + 3 \bigg(\K - \sfrac{\kappa\Bar{\rho} c^2}{2a} \bigg)\varphi = 4\pi G_{\!N}\, (\rho-\bar{\rho})\, .
\ee
For $\K \neq \kappa\Bar{\rho} c^2/(2a)$ we can rewrite this equation as
\be{3.3}
\Delta\phi -\nu\, \phi = 4\pi G_{\!N}\rho\, ,
\ee
where we introduced a new auxiliary function
\be{3.4} \phi = \varphi - \sfrac{4\pi G_{\!N}\, \bar{\rho}}{\nu} = \varphi - \sfrac{1}{3} c^2 a \bigg[ 1 - \sfrac{2\K}{3} \sfrac{|\Omega_\K|}{\Omega_{\rm M}}
\sfrac{a}{a_0} \bigg]^{-1} \ee
and a parameter
\be{3.5}
\nu \equiv 3\bigg(\sfrac{\kappa\bar{\rho}c^2}{2a}-\K\bigg)\neq 0\, .
\ee
Eq.~\rf{3.5} shows that the parameter $\nu$ is positive for the open or flat Universe, but changes sign from plus to minus with growth of the scale factor $a$ for
the closed Universe.

The mass density $\rho$ is given by Eq.~\rf{2.6}. Therefore, we can consider the total function $\phi$ as a superposition of individual functions $\phi_i$, each
corresponding to the $i$-th gravitating mass. Then the function $\phi_i$ satisfies the following equation outside the point-like source (located at the origin of
coordinates):
\be{3.6}
\Delta\phi_i -\nu\phi_i = 0 \, ,
\ee
which for the metric \rf{2.1} can be written in the form
\be{3.7}
\sfrac{1}{\Sigma^2(\chi)} \der{}{\chi} \bigg( \Sigma^2(\chi) \,\der{\phi_i}{\chi} \bigg) - \nu \phi_i = 0\, .
\ee
With the help of the definitions
\be{3.8}
U(\eta,\chi) \equiv \Sigma(\chi)\,\phi_i(\eta,\chi)
\ee
and
\be{3.9}
\mu \equiv \sfrac{1}{\Sigma(\chi)} \dder{\Sigma(\chi)}{\chi} + \nu =
\begin{cases}
	\nu-1 & \;\text{for}\; \K=+1 \\
	\nu & \;\text{for}\; \K=0 \\
	\nu+1 & \;\text{for}\; \K=-1 \\
\end{cases}
\ee
Eq.~\rf{3.7} can be presented in the following form:
\be{3.10}
\dder{U}{\chi} - \mu U = 0\, .
\ee
Then the general solution is
\be{3.11}
\begin{aligned}
	\phi_i &= \sfrac{A_1\sin\!\big(\sqrt{|\mu|}\chi\big) + A_2 \cos\!\big(\sqrt{|\mu|}\chi\big)}{\Sigma(\chi)} \;, \quad \mu < 0\, ; \\[2mm]
	\phi_i &= \sfrac{B_1\,\chi + B_2}{\Sigma(\chi)} \;, \hspace{4cm} \mu = 0\, ; \\[2mm]
	\phi_i &= \sfrac{C_1\,e^{-\sqrt{\mu}\chi} + C_2\,e^{\sqrt{\mu}\chi}}{\Sigma(\chi)} \;, \hspace{2.2cm} \mu > 0\, .
\end{aligned}
\ee

In addition, we introduce a new parameter
\ba{3.12}
&{}&  \lambda_{\rm phys}^{-1} \equiv \sfrac{\sqrt{|\mu|}}{a}\nn \\
&=& \sqrt{\sfrac{9 H_0^2\,\Omega_{\rm M}}{2 c^2} \big(z+1\big)^{\!3} \,\bigg[ 1 - \sfrac{8\K}{9} \sfrac{|\Omega_\K|}{\Omega_{\rm M}} \sfrac{1}{z+1} \bigg]}\, ,
\ea
where $z=(a_0/a)-1$ is the redshift. In what follows, this parameter will define a characteristic length of cosmic screening. It is worth noting that the
introduced screening length is a dynamical function since it depends on the scale factor $a$.

Let us investigate three curvature types separately.

\section{Flat Universe}
\label{Sec4}

\setcounter{equation}{0}

For the flat case $\mu=\nu > 0$. Then the solution \rf{3.11} is
\be{4.1}
\phi_i = \sfrac{C_1\,e^{-\sqrt{\mu}\chi} + C_2\,e^{\sqrt{\mu}\chi}}{\chi}\, .
\ee
Applying Newtonian limit $\phi_i(\chi \rightarrow 0) \rightarrow -G_{\!N} m_i /\chi$ and the boundary condition $\phi_i(\chi \rightarrow +\infty) \rightarrow 0$,
we get
\be{4.2} \phi_i = -\sfrac{G_{\!N} m_i}{r}\, e^{-\sqrt{\mu} r} \;, \qquad 0<r<+\infty\, , \ee
where in the flat Universe $\chi\equiv r$ is the absolute value of the three-dimensional comoving radius-vector. As usual, the physical radius-vector is defined
as $\textbf{r}_{\rm phys}=a\textbf{r}$.

In \rf{4.2} the origin of coordinates is located on the gravitating mass. For a many-particle system the total function $\varphi$ takes the form
\ba{4.3}
\varphi &=& \sum_i \phi_i + \sfrac{4\pi G_{\!N} \bar{\rho}}{\nu} \nn \\
&=& \sfrac{1}{3} c^2 a - G_{\!N} \sum_i \sfrac{m_i}{|\textbf{r}-\textbf{r}_i|}\, e^{-\sqrt{\mu}|\textbf{r}-\textbf{r}_i|}\, .
\ea
Therefore, for the total gravitational potential we get
\be{4.4} \Phi = \sfrac{1}{3} - \sfrac{G_{\!N}\,}{c^2 a} \sum_i \sfrac{m_i}{|\textbf{r}-\textbf{r}_i|}\, e^{-\sqrt{\mu}|\textbf{r}-\textbf{r}_i|} \ee
that exactly coincides with the result of \cite{Eingorn:2015hza} where it was also shown that the average value of \rf{4.4} is equal to zero: $\bar\Phi=0$, as it
should be for the first-order perturbations (see, e.g., the corresponding discussion in \cite{EBV}).

In the considered case, the screening length at the present time $a=a_0$ with the values \rf{2.5} is estimated, in accord with \cite{Eingorn:2015hza}, as
\be{4.5} \lambda_{\rm phys}^{\!(0)} =\frac{a_0}{\sqrt{\mu_0}}= \bigg(\sfrac{9 H_0^2\,\Omega_{\rm M}}{2 c^2} \bigg)^{\!\!-1/2} \approx 3736\,{\rm Mpc}\, . \ee

\section{Open Universe}
\label{Sec5}

\setcounter{equation}{0}

For the case $\K=-1$ the parameter $\nu$ is positive, therefore $\mu=\nu+1 > 0$. Then from Eq.~\rf{3.11} we get
\be{5.1}
\phi_i = \sfrac{C_1\,e^{-\sqrt{\mu}\chi} + C_2\,e^{\sqrt{\mu}\chi}}{\sinh\!\chi}\, .
\ee
Applying Newtonian limit $\phi_i(\chi \rightarrow 0) \rightarrow -G_{\!N} m_i /\chi$ and the boundary condition $\phi_i(\chi \rightarrow +\infty) \rightarrow 0$,
we obtain
\be{5.2} \phi_i = - \sfrac{G_{\!N} m_i}{\sinh\!\chi}\,e^{-\sqrt{\mu}\chi}  \;, \qquad 0<\chi<+\infty\, . \ee
Here the origin of coordinates is located on the gravitating mass. For a many-particle system the total function $\varphi$ takes the form
\be{5.3} \varphi = \sfrac{1}{3} c^2 a \bigg[ 1 + \sfrac{2}{3} \sfrac{\Omega_\K}{\Omega_{\rm M}} \sfrac{a}{a_0} \bigg]^{-1} - G_{\!N} \sum_i \sfrac{m_i}{\sinh
	l_i}\, e^{-\sqrt{\mu}l_i}\, , \ee
where $l_i$ denotes the geodesic distance between the $i$-th mass $m_i$ and the point of observation.
Therefore, the total gravitational potential is
\be{5.4}
\Phi = \sfrac{1}{3} \bigg[ 1 + \sfrac{2}{3} \sfrac{\Omega_\K}{\Omega_{\rm M}} \sfrac{a}{a_0} \bigg]^{-1} - \sfrac{G_{\!N}\,}{c^2 a}
\sum_i \sfrac{m_i}{\sinh\!l_i}\, e^{-\sqrt{\mu}l_i}\, .
\ee
Similarly to the flat Universe case, here we also expect that the average value of the potential \rf{5.4} is equal to zero. To prove it, we consider first the
average value of the individual contribution \rf{5.2}:
\ba{5.5}
\bar{\phi}_i &=& \sfrac{1}{V} \int_V \phi_i\,dV \nn\\
&=& -G_{\!N} m_i \sfrac{4\pi}{V} \int_0^{+\infty} \sfrac{1}{\sinh\!\chi}\, e^{-\sqrt{\mu}\chi}\; \sinh^2\!\chi\,d\chi\nn\\
&=& -\sfrac{1}{3} c^2 a \sfrac{m_i}{V} \sfrac{1}{\bar{\rho}} \bigg[ 1 + \sfrac{2}{3} \sfrac{\Omega_\K}{\Omega_{\rm M}} \sfrac{a}{a_0} \bigg]^{-1}\, . \ea
Then for the average value of the total gravitational potential we get
\be{5.6} \bar{\Phi} = \sfrac{1}{3} \bigg[ 1 + \sfrac{2}{3} \sfrac{\Omega_\K}{\Omega_{\rm M}} \sfrac{a}{a_0} \bigg]^{-1} + \sfrac{1}{c^2 a} \sum_i \bar{\phi_i}=0\,
, \ee
where we have taken into account that $\left(\sum_i m_i\right)/V=\bar\rho$.

For the open Universe, the screening length at the present time $a=a_0$ with the values \rf{2.5} is
\be{5.7}
\lambda_{\rm phys}^{\!(0)} = \bigg(\sfrac{9 H_0^2\,\Omega_{\rm M}}{2 c^2} \,\bigg[ 1+\sfrac{8}{9} \sfrac{\Omega_\K}{\Omega_{\rm M}} \bigg]\bigg)^{\!-1/2}
\approx 3732\,{\rm Mpc}\, .
\ee

\section{Closed Universe}
\label{Sec6}

\setcounter{equation}{0}

\begin{figure}[!ht]
	\centering
	\begin{tabular}{@{}c@{}}
		\includegraphics[scale=.25]{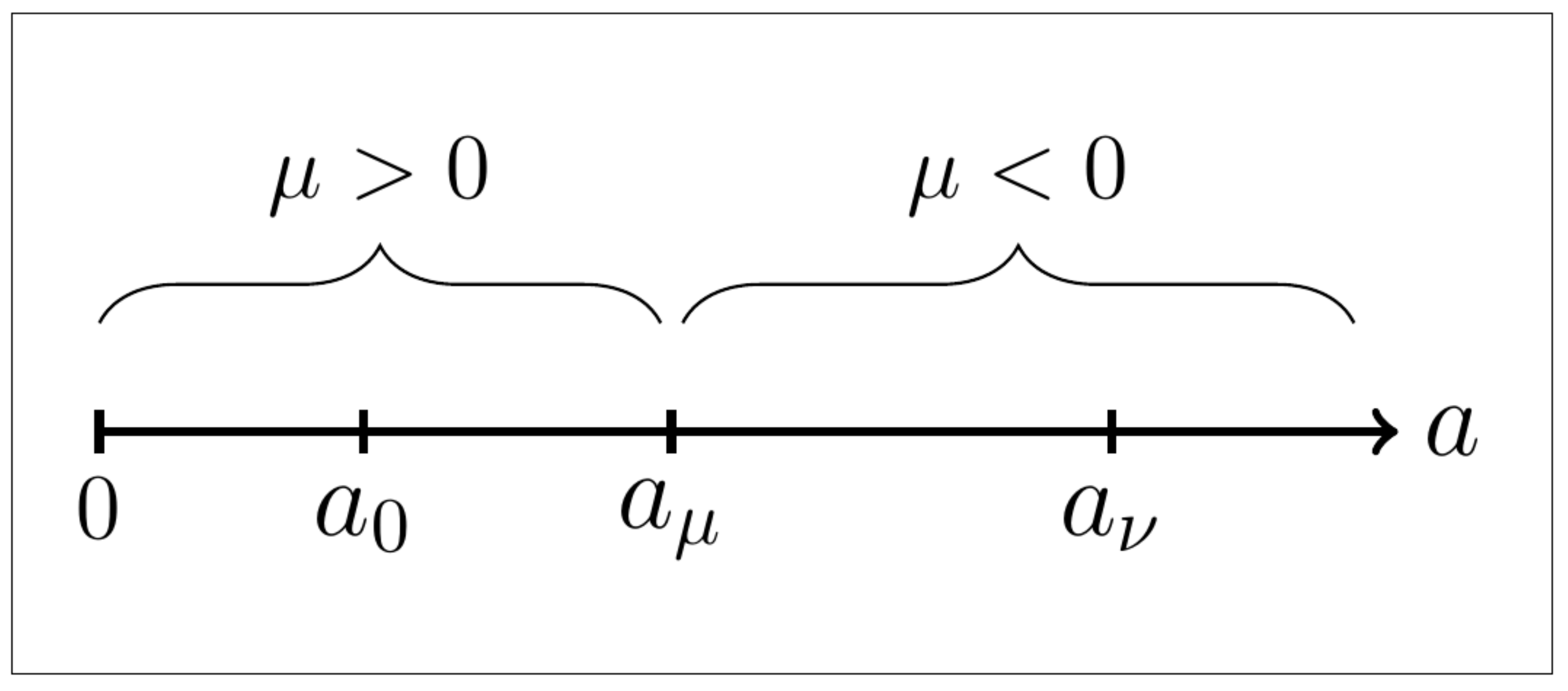}
	\end{tabular}
	\caption{Schematic illustration of the relation between the scale
		factor and the parameter $\mu$. The points $a_\mu$ and $a_\nu$ represent the cases $\mu=0$ and $\nu=0$ ($\mu=-1$), respectively; $a_0$ stands for the present
		value of the scale factor.}
	\label{fig:schematic}
\end{figure}

In the case of the closed Universe the parameter $\mu$ reads
\be{6.1}
\mu =\nu -1 = \sfrac{3\kappa\bar{\rho}c^2}{2a}-4=
\sfrac{3\kappa\bar{\rho}c^2}{2a}\bigg(1-\frac{8}{9}\frac{|\Omega_\K|}{\Omega_M}\frac{a}{a_0}\bigg)\, .
\ee
Therefore, with increasing scale factor $a$ from zero to infinity this parameter changes its sign from positive to negative and tends to $-4$ for $a\to +\infty$.
Hence, for the negative values of $\mu$ the only integer value of its square root is $\sqrt{|\mu|}=1$ (for finite values of the scale factor), that takes place
for $\mu=-1 \Leftrightarrow \nu=0$. Since $|\Omega_\K|\ll\Omega_M$ (see Eq.~\rf{2.5}), at the present time $a=a_0$ the parameter $\mu=\mu_0$ is positive:
$\mu_0>0$. From Eq.~\rf{6.1} we can also find two special values of the scale factor. The first one is
\be{6.2} \mu=0 \quad \Rightarrow \quad a_{\mu}= \frac{3\kappa\bar\rho c^2}{8}\, . \ee
At this value of $a$ the parameter $\mu$ changes its sign. The second one is
\be{6.3}
\mu=-1 \quad \Rightarrow \quad
a_{\nu}= \frac{\kappa\bar\rho c^2}{2}
\ee
and corresponds to the zero value of $\nu$. Obviously, for $\nu=0$ the transformation \rf{3.4} does not work. Schematic location of the special values \rf{6.2}
and \rf{6.3} of the scale factor is depicted in Fig.~\ref{fig:schematic}.

Now we turn to the solution for the gravitational potential starting from the case $\mu\neq-1 \Leftrightarrow \nu\neq 0$. Then from \rf{3.11} we get
\be{6.4}
\begin{aligned}
	\phi_i &= \sfrac{A_1\sin(\sqrt{|\mu|}\chi) + A_2 \cos(\sqrt{|\mu|}\chi)}{\sin\!\chi} \;, \\[1mm]
	&\hspace{3.5cm} (\mu < 0) \wedge (\mu \neq -1)\, ; \\[1mm]
	\phi_i &= \sfrac{B_1\,\chi + B_2}{\sin\!\chi} \;, \hspace{3.1cm} \mu = 0\, ; \\[2mm]
	\phi_i &= \sfrac{C_1\,e^{-\sqrt{\mu}\chi} + C_2\,e^{\sqrt{\mu}\chi}}{\sin\!\chi} \;, \hspace{1.3cm} \mu > 0\, .
\end{aligned}
\ee
First, we consider the solution with negative $\mu$. The condition of regularity of this solution at $\chi=\pi$ requires
\be{6.5}
A_1\sin(\sqrt{|\mu|}\pi) + A_2 \cos(\sqrt{|\mu|}\pi) = 0\, ,
\ee
which gives
\be{6.6}
A_1 = -A_2\,\sfrac{\cos(\sqrt{|\mu|}\pi)}{\sin(\sqrt{|\mu|}\pi)} \, ,\quad \mu\neq -1\, .
\ee
Therefore,
\be{6.7} \phi_i = A_2\, \sfrac{\sin\!\big[\sqrt{|\mu|}(\pi-\chi)\big]}{\sin\big(\sqrt{|\mu|}\pi\big) \sin\!\chi} \;. \ee
Now, if we employ the Newtonian limit $\phi_i(\chi \rightarrow 0) \rightarrow \phi^{(N)}_i=-G_{\!N} m_i /\chi$, we obtain
\be{6.8} \phi_i = -G_{\!N} m_i\, \sfrac{\sin\!\big[\sqrt{|\mu|}(\pi-\chi)\big]}{\sin\big(\sqrt{|\mu|}\pi\big) \sin\!\chi}\, . \ee

Following the same procedure for the cases $\mu=0$ and $\mu>0$, we get:
\be{6.9}
\begin{aligned}
	\phi_i &= -G_{\!N} m_i\, \sfrac{\sin\!\big[\sqrt{|\mu|}(\pi-\chi)\big]}{\sin\!\big(\sqrt{|\mu|}\pi\big) \sin\!\chi} \, ,\\[1mm]
	&{}\hspace{3.4cm} (\mu < 0) \wedge (\mu \neq -1)\, ; \\[1mm]
	\phi_i &= -G_{\!N} m_i\,\sfrac{\pi-\chi}{\pi \sin\!\chi} \;, \hspace{2.35cm} \mu = 0\, ; \\[2mm]
	\phi_i &= -G_{\!N} m_i\, \sfrac{\sinh\!\big[\sqrt{\mu}(\pi-\chi)\big]}{\sinh\!\big(\sqrt{\mu}\pi\big) \sin\!\chi} \, , \hspace{0.6cm} \mu > 0 \;.
\end{aligned}
\ee

Let us now consider the exceptional case $\mu=-1 \Leftrightarrow \nu=0$. Before that, it is worth noting that in the closed Universe, unlike the flat and open
Universe cases, we can determine the individual contribution of each mass into the total average comoving mass density:
\be{6.10} \bar{\rho} = \sum_i \sfrac{m_i}{V} \equiv \sum_i \bar{\rho}_i\, , \ee
where $V=2\pi^2$ is the comoving space volume. Then we can solve Eq.~\rf{3.2} for each combination $(m_i,\bar\rho_i)$ separately. For example, the function
$\varphi_i$ outside the $i$-th mass satisfies the equation
\be{6.11} \Delta\varphi_i = -4\pi G_{\!N} \bar{\rho}_i\, . \ee
The solution reads
\be{6.12} \varphi_i = A_i - G_{\!N} m_i \,\sfrac{\cos\!\chi}{\sin\!\chi} \bigg(\!1-\sfrac{\chi}{\pi}\bigg)\, , \ee
where $A_i$ is the constant of integration and the second constant has been determined by the demand for regularity of the potential at $\chi=\pi$. It can be
easily seen that the potential \rf{6.12} has the Newtonian limit for $\chi\to 0$. The constant $A_i$ can be found from the natural condition that the average
values of the first-order perturbations, i.e. the gravitational potential in our case, should be equal to zero:
\be{6.13}
\bar{\varphi} = \sum_i \sfrac{1}{V} \int_V \varphi_i\,dV =\sum_i\bar\varphi_i\, ,
\ee
\ba{6.14}
\bar{\varphi}_i &=& \sfrac{4\pi}{V} \int_0^{\pi} \bigg[ A_i - G_{\!N} m_i \,\sfrac{\cos\!\chi}{\sin\!\chi}
\bigg(\!1-\sfrac{\chi}{\pi}\bigg) \bigg] \sin^2\!\chi\,d\chi\nn\\
&=& \sfrac{2\pi}{V} \bigg[ A_i\pi - \sfrac{G_{\!N} m_i}{2} \bigg]=0\, . \ea
Thus, $A_i=G_{\!N} m_i/(2\pi)$. Consequently, the solution \rf{6.12} becomes
\be{6.15} \varphi_i = \sfrac{G_{\!N} m_i}{2\pi} - G_{\!N} m_i \,\sfrac{\cos\!\chi}{\sin\!\chi} \bigg(\!1-\sfrac{\chi}{\pi}\bigg)\, . \ee
Therefore, taking into account Eq.~\rf{6.10}, the complete set of solutions, including the exceptional one \rf{6.15}, is
\begin{subequations}
	\begin{align}
		\varphi_i &= -G_{\!N} m_i\, \bigg\{ \sfrac{\sin\!\big[\sqrt{|\mu|}(\pi-\chi)\big]}{\sin\!\big(\sqrt{|\mu|}\pi\big) \sin\!\chi} +
		\sfrac{2}{\pi(|\mu|-1)} \bigg\}\, , \nn\\
		&{}\hspace{3.2cm} (\mu < 0) \wedge (\mu \neq -1)\, ; \label{6.16a} \\[2mm]
		\varphi_i &= -G_{\!N} m_i\, \bigg\{ \sfrac{\cos\!\chi}{\sin\!\chi} \bigg(\!1-\sfrac{\chi}{\pi}\bigg) - \sfrac{1}{2\pi} \bigg\} \;,
		\hspace{0.3cm} \mu = -\, 1; \label{6.16b} \\[2mm]
		\varphi_i &= -G_{\!N} m_i\, \bigg\{ \sfrac{1}{\sin\!\chi} \bigg(\!1-\sfrac{\chi}{\pi}\bigg) - \sfrac{2}{\pi} \bigg\} \;,
		\hspace{0.6cm} \mu = 0\, ; \label{6.16c} \\[2mm]
		\varphi_i &= -G_{\!N} m_i\, \bigg\{ \sfrac{\sinh\!\big[\sqrt{\mu}(\pi-\chi)\big]}{\sinh\!\big(\sqrt{\mu}\pi\big) \sin\!\chi} -
		\sfrac{2}{\pi(\mu+1)} \bigg\} \;,\nn\\
		&{}\hspace{4.2cm} \mu > 0\, . \label{6.16d}
	\end{align}
	\label{6.16}%
\end{subequations}

The average values of all these individual potentials are equal to zero. We have already discussed this fact with respect to \rf{6.16b}. The same can be proven
for the other expressions. For example, for \rf{6.16a} we have
\ba{6.17}
&{}&-\frac{\bar{\varphi}_i}{G_{\!N} m_i} = -\sfrac{1}{G_{\!N} m_i V} \int_V \varphi_i\,dV \nn\\
&=& \sfrac{4\pi}{V} \int_0^{\pi} \bigg\{ \sfrac{\sin\!\big[u(\pi-\chi)\big]}{\sin\!\big(u\pi\big) \sin\!\chi} + \sfrac{2}{\pi(u^2-1)} \bigg\}
\sin^2\!\chi\,d\chi \nn\\
&=&
\sfrac{4\pi}{V} \bigg\{ \sfrac{1}{\sin\!\big(u\pi\big)} \sfrac{1}{1-u^2} \sin\!\big(u\pi\big) + \sfrac{1}{\pi(u^2-1)} \pi \bigg\} = 0,\nn\\
&{}&
\ea
where $u\equiv \sqrt{|\mu|}$.  Therefore, the average values of total gravitational potentials are
also equal to zero: $\bar\Phi =(c^2 a)^{-1} \sum_i\bar\varphi_i =0$.

The first derivatives of the potentials \rf{6.16} with respect to $\chi$ define the gravitational force $F_i\equiv-\partial\varphi_i/\partial\chi$ (per unit mass
and up to the prefactor $1/a^2$) induced by the mass $m_i$. From \rf{6.16a}-\rf{6.16d} we get, respectively,
\begin{subequations}
	\begin{align}
		&F_i =- G_{\!N} m_i\, \left\{\sfrac{\sqrt{|\mu|} \cos\!\big[\sqrt{|\mu|}(\pi-\chi)\big]}{{\sin\!\chi
				\sin\!\big(\sqrt{|\mu|}\pi\big)}}\right.\nn\\
		&+ \left.\frac{\cos\!\chi \sin\!\big[\sqrt{|\mu|}(\pi-\chi)\big]}{\sin^2\!\chi \sin\!\big(\sqrt{|\mu|}\pi\big)}\right\} \;,
		{}\hspace{.6cm} (\mu < 0) \wedge (\mu \neq -1)\, ; \label{6.19a} \\[2mm]
		&F_i = -G_{\!N} m_i\,\bigg[\sfrac{\cos\!\chi}{\pi \sin\!\chi} + \sfrac{1}{\sin^2\!\chi}\bigg(1-\sfrac{\chi}{\pi}\bigg)\bigg] \;,
		\hspace{0.3cm} \mu = -1; \label{6.19b} \\[2mm]
		&F_i =- G_{\!N} m_i\,\bigg[ \sfrac{1}{\pi \sin\!\chi} + \sfrac{\cos\!\chi}{ \sin^2\!\chi}\bigg(1-\sfrac{\chi}{\pi}\bigg)  \bigg] \;,
		\hspace{0.6cm} \mu = 0; \label{6.19c} \\[2mm]
		&F_i = -G_{\!N} m_i\, \left\{\sfrac{\sqrt{\mu} \cosh\!\big[\sqrt{\mu}(\pi-\chi)\big]}{\sin\!\chi \sinh\!\big(\sqrt{\mu}\pi\big)}\right.\nn\\
		&+\left.\frac{\cos\!\chi \sinh\!\big[\sqrt{\mu}(\pi-\chi)\big]}{\sin^2\!\chi \sinh\!\big(\sqrt{\mu}\pi\big)}\right\} \;, {}\hspace{.5cm} \mu > 0
		\label{6.19d} \;.
	\end{align}
	\label{6.19}
\end{subequations}

\begin{figure}[!ht]
	\centering
	\begin{tabular}{@{}c@{}}
		\includegraphics[scale=.4]{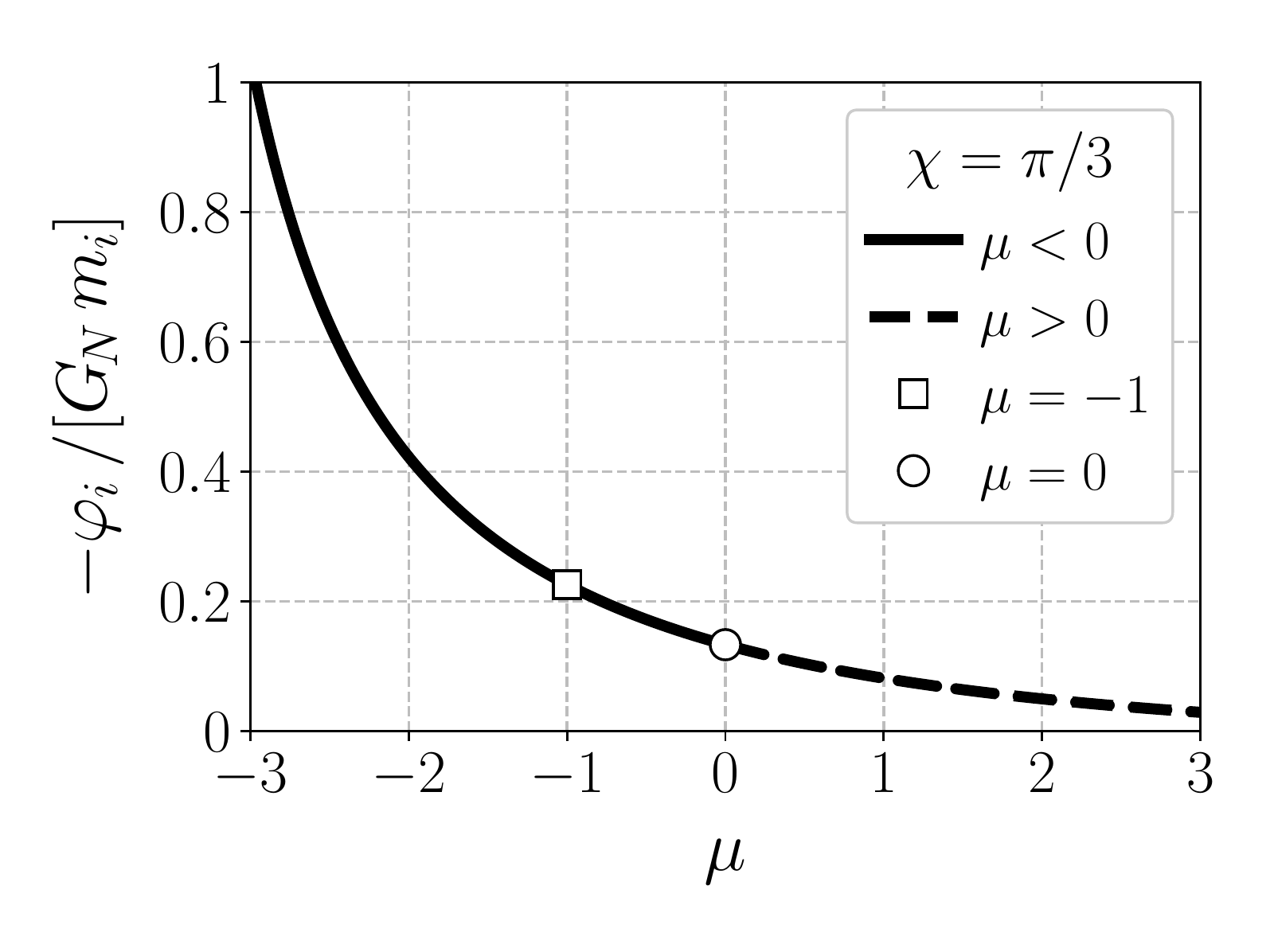}
	\end{tabular}
	\vspace{-2mm}
	\caption{Plot of Eq.~\rf{6.16} with fixed $\chi=\pi/3$.}
	\vspace{4mm}
	\label{fig:mr_zero}
\end{figure}

It is not difficult to verify that the solutions \rf{6.16a}-\rf{6.16d} are smoothly connected with each other: these functions as well as their first derivatives
\rf{6.19a}-\rf{6.19d} are matched at $\mu=-1$ and $\mu=0$ for any value of $\chi \in (0,\pi]$. We demonstrate this graphically in Fig.~\ref{fig:mr_zero} where we
put for definiteness $\chi=\pi/3$.

Taking into account that $\mu \in (-4,+\infty)$ and the limiting value $-4$ does not correspond to any finite value of the scale factor $a$, it is not difficult
to verify that the first derivatives \rf{6.19} are equal to zero only at the antipodal point $\chi=\pi$. Moreover, the second derivatives of the functions
\rf{6.16} are negative at $\chi=\pi$. Therefore, the potentials \rf{6.16} represent monotonically increasing functions from $-\infty$ (for $\chi \to 0$) to the
following positive maximal values (at $\chi=\pi$):
\begin{subequations}
	\begin{align}
		\varphi_i(\chi\rightarrow\pi) &= -G_{\!N} m_i\, \bigg\{ \sfrac{\sqrt{|\mu|}}{\sin\!\big(\sqrt{|\mu|}\pi\big)} + \sfrac{2}{\pi(|\mu|-1)} \bigg\} \;,
		\nn\\ &{}\hspace{2.5cm} (\mu < 0) \wedge (\mu \neq -1); \label{6.20a} \\[2mm]
		\varphi_i(\chi\rightarrow\pi) &= \sfrac{3G_{\!N} m_i}{2\pi} \;, \quad \mu = -1; \label{6.20b} \\[2mm]
		\varphi_i(\chi\rightarrow\pi) &= \sfrac{G_{\!N} m_i}{\pi} \;, \quad\ \,\mu = 0; \label{6.20c} \\[2mm]
		\varphi_i(\chi\rightarrow\pi) &= -G_{\!N} m_i\, \bigg\{ \sfrac{\sqrt{\mu}}{\sinh\!\big(\sqrt{\mu}\pi\big)} - \sfrac{2}{\pi(\mu+1)} \bigg\} \;,\nn\\
		&{}\hspace{4cm} \mu > 0 \label{6.20d} \;.
	\end{align}
	\label{6.20}%
\end{subequations}
Obviously, the limiting values \rf{6.20} are matched at $\mu=-1$ and $\mu=0$.

Since the first derivatives of the potentials \rf{6.16} are equal to zero at $\chi=\pi$, the gravitational force induced by the $i$-th mass $m_i$ is equal to zero
at the antipodal point with respect to this mass. This is an interesting feature of the closed Universe.

To conclude this section, we compare the behavior of the potential $\varphi_i$ and the corresponding force $F_i$ with the Newtonian expressions at the present
time $a=a_0$. At this moment the parameter $\mu=\mu_0$ and the screening length $\lambda_{\rm phys}^{\!(0)}=a_0/\sqrt{\mu_0}$ are:
\be{6.21} \mu_0= \sfrac{9}{2}\sfrac{\Omega_{\rm M}}{|\Omega_\K|}-4 \approx 28.22\, , \ee
\be{6.22}
\lambda_{\rm phys}^{\!(0)} = \bigg(\sfrac{9 H_0^2\,\Omega_{\rm M}}{2 c^2} \,\bigg[ 1-\sfrac{8}{9} \sfrac{|\Omega_\K|}{\Omega_{\rm M}} \bigg]\bigg)^{\!-1/2}
\approx 3992\,{\rm Mpc}\, ,
\ee
where we have used the values of the cosmological parameters given in \rf{2.5}.

In the case of positive $\mu$, the expressions for the gravitational potential and force are given by the formulas \rf{6.16d} and \rf{6.19d}, respectively. At the
moment $a=a_0$ the potential \rf{6.16d} can be written in the form
\ba{6.23} \varphi_i^{(0)} &=& -G_{\!N} m_i \left\{ \sfrac{\sinh\!\Big(\!\sqrt{\mu_0}\pi - \tilde{r}_{\rm phys}\Big)}{\sinh\!\Big(\!\sqrt{\mu_0}\pi\Big)
	\sin\!\Big(\tilde{r}_{\rm phys} \big/ \sqrt{\mu_0} \Big)}\right. \nn\\
&-&\left. \sfrac{2}{\pi\big(\mu_0+1\big)}\right\} \;,
\ea
where $\varphi_i^{(0)}\equiv \varphi_i\Big|_{a=a_0}$ and we have introduced the dimensionless physical distance
\be{6.24} \tilde{r}_{\rm phys} \equiv \frac{r^{(0)}_{\rm phys}}{\lambda_{\rm phys}^{\!(0)}}=\frac{a_0\chi}{\lambda_{\rm phys}^{\!(0)}} = \chi\sqrt{\mu_0} \in
\big[0,\sqrt{\mu_0}\pi\big]\, . \ee
The Newtonian potential can be presented in the form
\be{6.25} \varphi_i^{\scriptscriptstyle{(N)}} =-\frac{G_{\!N} m_i}{\chi}=- \frac{G_{\!N} m_ia_0}{r^{(0)}_{\rm phys}} =-\frac{G_{\!N}
	m_i\sqrt{\mu_0}}{\tilde{r}_{\rm phys}}\, . \ee
The dimensionless form of these potentials is:
\be{6.26} \tilde{\varphi}_i^{(0)} \equiv \sfrac{1}{\sqrt{\mu_0}} \sfrac{\varphi_i^{(0)}}{G_{\!N} m_i}\, , \ee
\be{6.27}
\tilde{\varphi}_i^{\scriptscriptstyle{(N)}} \equiv \sfrac{1}{\sqrt{\mu_0}} \sfrac{\varphi_i^{\scriptscriptstyle{(N)}}}{G_{\!N} m_i} =
-\sfrac{1}{\tilde{r}_{\rm phys}}\, .
\ee
Similarly, we can introduce the dimensionless analog of the gravitational force \rf{6.19d},
\ba{6.28}
&{}&\tilde{F}_i^{(0)} \equiv \sfrac{1}{\mu_0} \sfrac{F_i^{(0)}}{G_{\!N} m_i}\nn\\
&=&-\sfrac{\sin\left(\tilde{r}_{\rm phys}\big/\sqrt{\mu_0}\right) \cosh\!\big[\sqrt{\mu_0}\pi-\tilde{r}_{\rm phys}\big]}{\sqrt{\mu_0} \sin^2
	\left(\tilde{r}_{\rm phys}\big/\sqrt{\mu_0}\right) \sinh\!\big(\sqrt{\mu_0}\pi\big)}\nn\\
&-&\frac{\cos \left(\tilde{r}_{\rm phys}\big/\sqrt{\mu_0}\right) \sinh\!\big[\sqrt{\mu_0}\pi-\tilde{r}_{\rm phys}\big]}{\mu_0\sin^2 \left(\tilde{r}_{\rm
		phys}\big/\sqrt{\mu_0}\right) \sinh\!\big(\sqrt{\mu_0}\pi\big)}\, , \ea
and the dimensionless expression for Newtonian force,
\be{6.29}
\tilde{F}_i^{\scriptscriptstyle{(N)}} \equiv \sfrac{1}{\mu_0} \sfrac{F_i^{\scriptscriptstyle{(N)}}}{G_{\!N} m_i} =  -\sfrac{1}{\tilde{r}_{\rm phys}^2}\, .
\ee
The dimensionless potentials \rf{6.26}, \rf{6.27} and forces \rf{6.28}, \rf{6.29} are depicted in Fig.~\rf{fig:pot_closed}. The top picture shows a faster rush to the x-axis of the potential in the closed Universe compared to the Newtonian potential. Moreover, in contrast to the latter, the potential
$\tilde{\varphi}_i^{(0)}$ changes its sign crossing the x-axis. This is a necessary condition for the zero average value of the potential \rf{6.23}. In addition, $X\!=\!\left[\sinh\big(\!\sqrt{\mu_0}\pi\big)\right]^{-1}-2\left[\sqrt{\mu_0}\big(\mu_0+1\big)\pi\right]^{-1}\approx-4\times10^{-3}$ is the limiting value of $-\tilde{\varphi}_i^{(0)}$ for $\chi \to \pi$ (see Eq.~\rf{6.20d}). The force plots (bottom picture) demonstrate a faster drop in the absolute values of the force compared to the Newtonian expression. At the finite distance $\chi=\pi$ (i.e. at the antipodal point) the gravitational force is equal to zero.

\begin{figure}[!ht]
	\centering
	\begin{tabular}{@{}c@{}}
		\includegraphics[scale=.4]{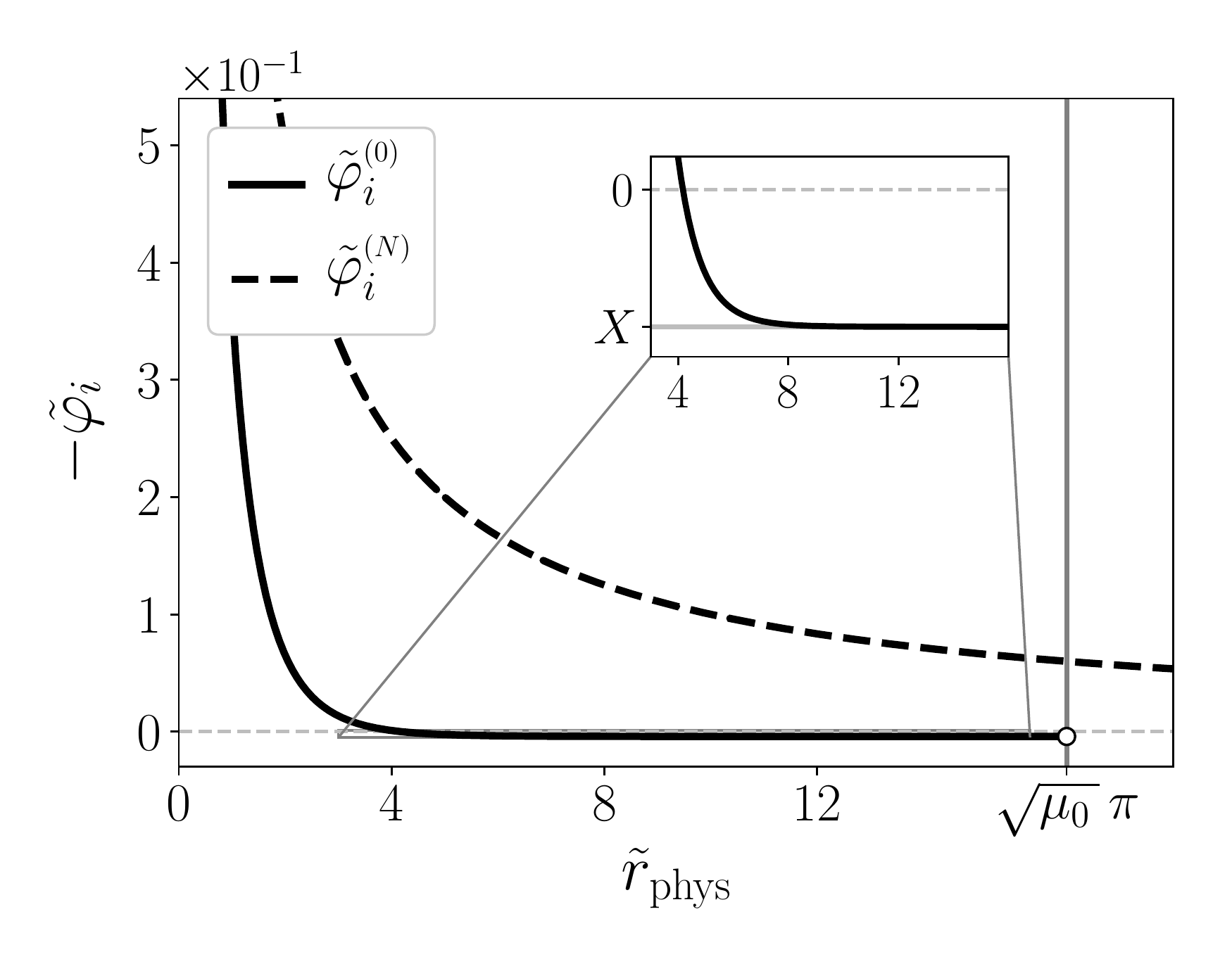}
	\end{tabular}
	\begin{tabular}{@{}c@{}}
		\includegraphics[scale=.4]{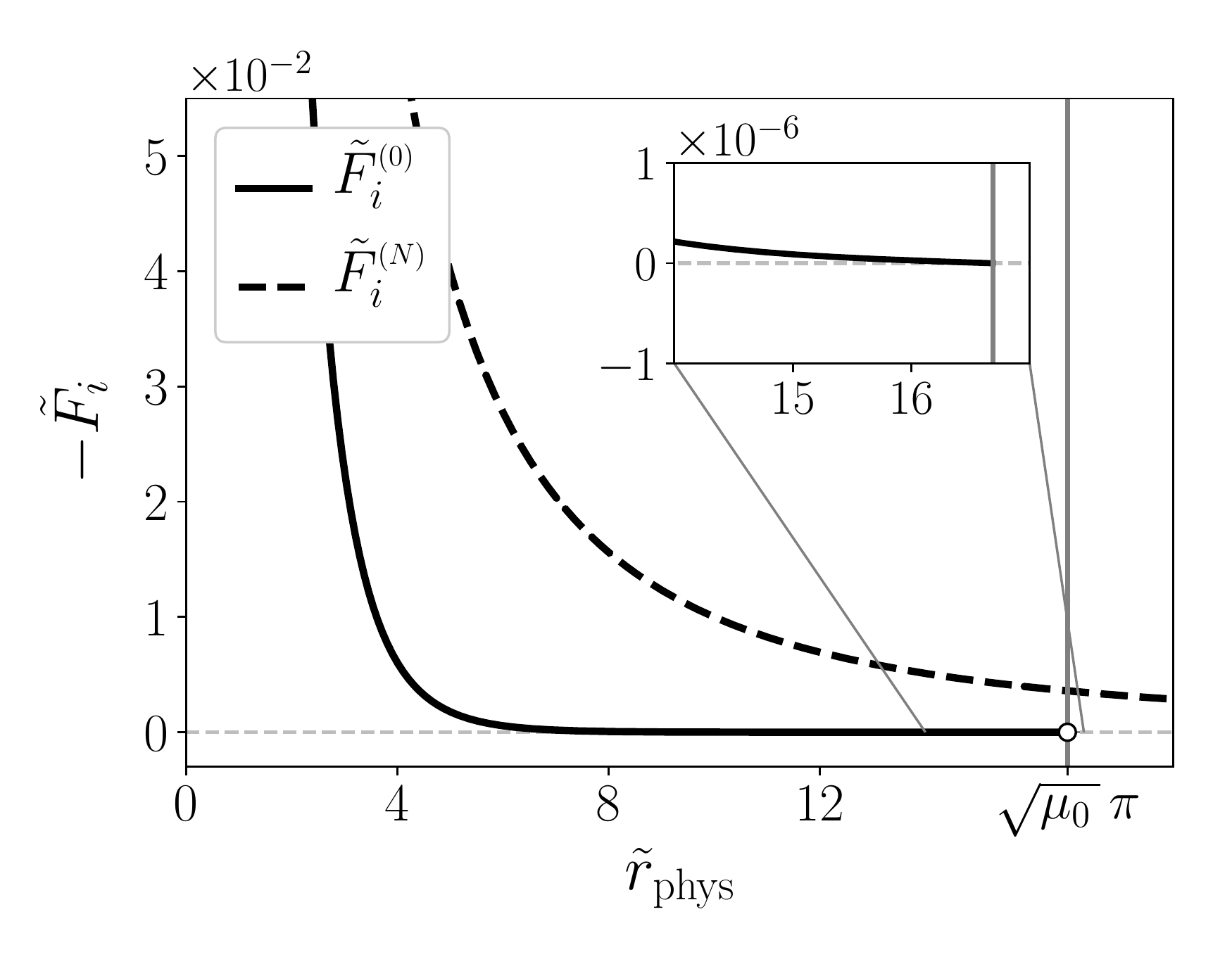}
	\end{tabular}
	\vspace{-2mm}
	\caption{Graphical representations of the gravitational potentials (top) defined by Eqs.~\rf{6.26}, \rf{6.27} and
		forces (bottom) defined by Eqs.~\rf{6.28}, \rf{6.29}.}
	\label{fig:pot_closed}
\end{figure}

\section{Conclusion}

\setcounter{equation}{0}

In this paper we have considered the effect of spatial curvature on the form of the gravitational potential produced by discrete massive sources in the open and
closed Universe cases. Within the cosmic screening approach this potential satisfies the Helmholtz-type equation where the Laplace operator is defined by the
metric of the constant curvature spaces. We have not included the peculiar velocities of discrete masses since they negligibly contribute to the potential
\cite{Eingorn:2015hza}. We have solved this equation exactly for the open and closed Universe cases. The flat Universe was considered earlier in the paper
\cite{Eingorn:2015hza} where it was shown that the gravitational potential undergoes the Yukawa-type exponential screening at cosmological scales. In the present paper we have shown that the spatial curvature of the Universe considerably affect the shape of the gravitational potential. Although in the open Universe we also observe the exponential screening, there is a prefactor $1/\sinh l$ (with $l$ denoting the geodesic distance between the mass and the point of observation) instead of $1/r$ as in the flat space. In the closed Universe the situation is even more complicated and interesting. First of all, the form of the potential depends on the sign of the time-dependent parameter $\mu$ \rf{6.1}. This parameter changes its sign from positive to negative with the growth of the scale factor $a$. Consequently, the potential changes its form with the growth of $a$ (see Eqs.~\rf{6.16a}-\rf{6.16d}). Second, we do not observe here the exponential damping of the potential. Instead, the potential produced by an individual mass grows with distance from $-\infty$ and reaches its positive maximal value at the antipodal point. At the same time, the gravitational force is equal to zero at this point (see Fig.~\rf{fig:pot_closed}).

We have also demonstrated that, similarly to the flat space \cite{Eingorn:2015hza}, in the open and closed Universe cases the average values of the total gravitational potentials are equal to zero, as it should be for the first-order perturbations. Formulas for the potentials and forces derived in the present paper can be used for investigations of motion of astrophysical objects (e.g., galaxies) in the open and closed Universes, and for simulations of the large scale structure formation. These formulas form the basis for the subsequent analysis of the second-order perturbations in the closed and open Universes (see, e.g., \cite{Eingorn-second,Duygu} for the flat case).



%


\end{document}